\documentclass{article}
\usepackage{amsmath}
\usepackage{bbm}
\usepackage{amsfonts}
\usepackage{dsfont}
\usepackage{amssymb}
\usepackage{mathrsfs}
\usepackage{calrsfs}
\usepackage{color}
\usepackage{epsfig}
\usepackage{graphicx}

\usepackage[super,square,comma,sort&compress]{natbib}
\usepackage[pagebackref,colorlinks,linkcolor=blue, citecolor=blue, hyperindex,pdfstartview=FitH,plainpages=false,CJKbookmarks]{hyperref}

\def\bc{\begin{center}}

\def\ec{\end{center}}
\def\be{\begin{equation}}
\def\ee{\end{equation}}
\def\ben{\begin{eqnarray}}
\def\een{\end{eqnarray}}
\def\bde{\begin{definition}}
\def\ede{\end{definition}}
\def\bpro{\begin{proof}}
\def\epro{\end{proof}}
\def\bth{\begin{theorem}}
\def\eth{\end{theorem}}
\def\ble{\begin{lemma}}
\def\ele{\end{lemma}}
\def\bpr{\begin{proposition}}
\def\epr{\end{proposition}}
\def\bco{\begin{corollary}}
\def\eco{\end{corollary}}
\numberwithin{equation}{subsection}

\newcommand{\omits}[1]{}

\definecolor{dyellow}{rgb}{1.,0.8,.0}
\definecolor{myblue}{rgb}{.1,.1,.7}
\definecolor{dcyan}{rgb}{.0,.6,.6}
\definecolor{dmagenta}{rgb}{0.6,0.0,0.6}
\definecolor{brown}{rgb}{0.6,0.2,0.}
\definecolor{darkblue}{rgb}{.0,.0,0.5}
\definecolor{darkred}{rgb}{0.75,0.0,0.0}
\definecolor{orange}{rgb}{1.,.6,.0}
\definecolor{dorange}{rgb}{0.8,.4,.0}
\definecolor{darkgreen}{rgb}{0.0,0.6,0.0}
\definecolor{purple}{rgb}{.4,.0,.4}

\renewcommand{\abstract}[1]{\par\noindent{\small{\bf Abstract\/}: #1}}
\newenvironment{proof}[1][Proof.]{\begin{trivlist}
\item[\hskip \labelsep {\bfseries #1}]}{\end{trivlist}}

\newcommand{\qed}{\nobreak \ifvmode \relax \else
     \ifdim\lastskip<1.5em \hskip-\lastskip
      \hskip1.5em plus0em minus0.5em \fi \nobreak
      \vrule height0.75em width0.5em depth0.25em\fi}
\newtheorem{theorem}{Theorem}[subsection]
\newtheorem{lemma}[theorem]{Lemma}
\newtheorem{definition}[theorem]{Definition}
\newtheorem{proposition}[theorem]{Proposition}
\newtheorem{corollary}[theorem]{Corollary}

\newcommand\btd{\raise 2pt
\hbox{$\hat\bigtriangledown$}\hskip 1.5pt}
\newcommand\bt{\raise 2pt
\hbox{$\bigtriangledown$}\hskip 1.5pt}

\numberwithin{equation}{section}
\begin{document}
\title{Bilinear B$\ddot{a}$cklund transformations and Lax pair for the Boussinesq equation
}
\date{  }
\author{
\small{Yong-Qiang Bai $^{a,b
}$,
Yan-Jun LV$^{b}$}\\
\small{$^{a}$ Institute of Contemporary Mathematics, Henan
University, Kaifeng
475004, P.R.China }\\
\small{$^{b}$School of Mathematics and Statistics,
Henan University, Kaifeng 475004, P.R.China}\\
}
\maketitle
\begin{abstract}
{Hirota's bilinear approach is a very effective method to construct
 solutions for soliton systems.  In terms of this method, the nonlinear equations can be transformed into linear equations, and can be solved by using perturbation method.  In this paper, we  study the bilinear Boussinesq equation and obtain its bilinear B\"{a}cklund transformation. Starting from this bilinear B\"{a}cklund transformation, we also derive its Lax pair and test its integrability.}
\end{abstract}
\begin{flushleft}
\textbf{Keywords:}  Hirota method, $D$-operator, $B\ddot{a}cklund$ transformation, Lax pair \\
\textbf{PACS:} 02.30.Ik, 02.40.Ma, 02.40.Gh
\end{flushleft}
\tableofcontents
\section{Introduction}
 The integrability and the searching for explict solution of nonlinear equations are always important and forefront research topics\cite{6,7,8,9}. Hirota¡¯s bilinear approach is an effective method to construct  solutions for soliton systems by now\cite{1,car,Mc}. Meanwhile, the Lax pair and B\"{a}cklund transformation are also very important in discussing nonlinear evolution equations\cite{12,13,16,18,19}. Similar to nonlinear partial differential equations, one can also use the appropriate method to obtain the bilinear B\"{a}cklund transformation, the Lax pair,  the Miura transformation and so forth for bilinear equations\cite{1,10,14,15}.

 The purpose of the present paper is to present our results on the bilinear approach to
the Boussinesq equation. We will obtain its bilinear B\"{a}cklund transformation and Lax pair.

The paper is organized as follows. In the next section, we will obtain  a new bilinear equation by eliminating a variable from the original Boussinesq equation, and the correspongding nonlinear partial differential equation. In section 3, we construct a bilinear B\"{a}cklund transformation for the bilinear  Boussinesq equation system. In section 4, we will be
devoted to the construction of lax pair for the equation. Final section contains our discussion and
conclusion.
\section{New equations from the Boussinesq equation system}
In this section, we start from the bilinear Boussinesq equation system to botain a new bilinear equation. By means of appropriate transformation and the properties of D operator, we also get a new nonlinear differential equation.
The Boussinesq equation system can be written as
\begin{equation}\label{eq:zps}
\left\{\begin{array}{ll}
(D_{1}^{4}+{3}D_{2}^{2})f\cdot f=0,\\
(D_{1}D_{m+3}-{3}D_{1}^{2}D_{2}D_{m}-\frac{1}{4}D_{4}D_m)f\cdot f=0,
\end{array} \right.
\end{equation}
where $ m\neq{3}k,\quad k\in\mathbb{Z}^{+},\quad D_k D_m=D_{t_k}D_{t_{m}}$.

From Eq. (\ref{eq:zps}), one can get the system of Boussinesq equations
\begin{equation}\label{eq:aaps}
\left\{\begin{array}{ll}
(D_x^{4}+{3}D_{t_{2}}^{2})f\cdot f=0,\\
(D_{x}D_{t_4}-{4}D_{x}^{3}D_{t_{2}})f\cdot f=0.
\end{array} \right.
\end{equation}

By taking
\begin{equation}
    g=f_{t_2},\qquad h=f_{t_2 t_2}.
\end{equation}
and using the following formula
\begin{equation}
\begin{array}{ll}
D_{t_{2}}^{2}f\cdot f={2}(f_{t_{2}t_{2}}f-f_{t_{2}}f)={2}(h f-g f),\\
D_{x}^{3}D_{t_{2}}f\cdot f={2}g_{xxx}f-{6}g_{xx}f_{x}+{6}g_x f_{xx}-{2}f_{xxx}g=2 D_x^{3}g\cdot f,
\end{array}
\end{equation}
we can eliminate $t_{2}$ from Eq.(\ref{eq:aaps}) and get the bilinear differential equation
\begin{equation}\label{eq:aaaps}
\left\{\begin{array}{ll}
D_x^{4}f\cdot f+{6}(h f-g f)=0,\\
D_{x}D_{t_4}f\cdot f-{8}D_x^{3}g\cdot f=0,\\
D_{x}D_{t_4}g\cdot f-{4}D_x^{3}h\cdot f=0.
\end{array} \right.
\end{equation}

If we take
\begin{equation}
    u={2}(\ln f)_{xx},\quad v=\frac{g}{f},\quad w=\frac{h}{f},
\end{equation}
and take advantage of the following property of $D-$operator
\begin{equation}
\begin{array}{ll}
\frac{D_x^{4}f\cdot f}{f^{2}}=u_{xx}+{3}u^{2},\\
\frac{D_xD_{t_4}f\cdot f}{f^{2}}={2}(\ln f)_{x t_{4}},\\
\frac{D_x^{3}g\cdot f}{f^{2}}=v_{xxx}+{3}u v_x,\\
\frac{D_{t_{4}}D_x g\cdot f}{f^{2}}=v_{t_{4}x}+{2}v(\ln f)_{x{t_{4}}},\\
\frac{D_x^{3}h\cdot f}{f^{2}}=w_{xxx}+{3}u w_x.
\end{array}
\end{equation}
We can obtain the nonlinear form of Eqs. (\ref{eq:aaaps})
\begin{equation}
\left\{\begin{array}{ll}
u_{xx}+{3}u^{2}+{6}(w-v)=0,\\
\partial^{-1}u_{t_{4}}-{8}(v_{xxx}+{3}u v_x)=0,\\
v_{x t_{4}}+v\partial^{-1}u_{t_{4}}-{4}(w_{xxx}+{3}u w_x)=0.
\end{array} \right.
\end{equation}
After eliminating $\partial^{-1}u_{t_{4}}$ from the above equations, one can get the nonlinear evolution equation
\begin{equation}
\left\{\begin{array}{ll}
u_{xx}+{3}u^{2}+{6}(w-v)=0,\\
v_{x t_{4}}+{8}v(v_{xxx}+{3}u v_x)-{4}(w_{xxx}+{3}u w_x)=0.
\end{array} \right.
\end{equation}

\section{Bilinear B\"{a}cklund Transformation}
In this section, we will discuss the bilinear B\"{a}cklund transformation of the Boussinesq equations
\begin{equation}\label{eq:ccps}
\left\{\begin{array}{ll}
(D_x^{4}+{3}D_{t_{2}}^{2})f\cdot f=0,\\
(D_{x}D_{t_4}-{4}D_{x}^{3}D_{t_{2}})f\cdot f=0.
\end{array} \right.
\end{equation}

 To begin with, we consider the first equation of the Eqs. (\ref{eq:ccps}). By means of the general method to obtain bilinear B\"{a}cklund transformation , one have
\begin{equation}\label{eq:cccps}
\begin{array}{ll}
P_{1}=[(D_x^{4}+{3}D_{t_{2}}^{2})f\cdot f]g^{2}-f^{2}[(D_x^{4}+{3}D_{t_{2}}^{2})g\cdot g]\\
\quad=[(D_x^{4}f\cdot f)g^{2}-f^{2}(D_x^{4}g\cdot g)]+{3}[(D_{t_{2}}f\cdot f)g^{2}-f^{2}(D_{t_{2}}g\cdot g)]\\
\quad={2}D_x(D_x^{3}f\cdot g)(g f)+{6}D_x(D_x^{2}f\cdot g)(D_xg\cdot f)+{6}D_{t_{2}}(D_{t_{2}}f\cdot g)(g f)\\
\quad={6}D_{t_{2}}[(D_{t_{2}}+a D_x^{2})f\cdot g](g f)-{6}a D_{t_{2}}(D_x^{2}f\cdot g)(g f)\\
\qquad\qquad+{6}D_x(D_x^{2}f\cdot g)(D_xg\cdot f)+{2}D_x(D_x^{3}f\cdot g)(g f).\\
\end{array}
\end{equation}
According to
\begin{equation}
  D_{t_{2}}(D_x^{2}f\cdot g)(g f)=D_x(D_{t_{2}}D_xf\cdot g)(g f)-D_x(D_{t_{2}}f\cdot g)(D_x g\cdot f),
\end{equation}
Eq. (\ref{eq:cccps}) become
\begin{equation}\label{eq:ccccps}
\begin{array}{ll}
P_{1}={6}D_{t_{2}}[(D_{t_{2}}+a D_x^{2})f\cdot g](g f)-{6}a[D_x(D_{t_{2}}D_xf\cdot g)(g f)-D_x(D_{t_{2}}f\cdot g)(D_x g\cdot f)]\\
\qquad\qquad\qquad\qquad+{6}D_x(D_x^{2}f\cdot g)(D_xg\cdot f)+{2}D_x(D_x^{3}f\cdot g)(g f)\\
\quad={6}D_{t_{2}}[(D_{t_{2}}+a D_x^{2})f\cdot g](g f)-{6}aD_x(D_{t_{2}}D_xf\cdot g)(g f)+{6}aD_x(D_{t_{2}}f\cdot g)(D_x g\cdot f)]\\
\qquad\qquad\qquad\qquad+{6}D_x(D_x^{2}f\cdot g)(D_xg\cdot f)+{2}D_x(D_x^{3}f\cdot g)(g f)\\
\quad={6}D_{t_{2}}[(D_{t_{2}}+a D_x^{2})f\cdot g](g f)+{6}D_x[(D_x^{2}+a D_{t_{2}})f\cdot g](D_x g\cdot f)\\
\qquad\qquad\qquad\qquad+{2}D_x[(D_x^{3}-{3}a D_{t_{2}}D_x)f\cdot g](g f).
\end{array}
\end{equation}
We can deduce from the above equation that it is necessary that $P_{1}=0$.\\
It is obvious that as $a=\pm1$, $(D_{t_{2}}+a D_x^{2})f\cdot g=0$ is equivalent to $(D_x^{2}+a D_{t_{2}})f\cdot g=0$.\\
Thereafter, we can get
\begin{equation}\label{eq:ddps}
\left\{\begin{array}{ll}
(D_{t_{2}}+a D_x^{2})f\cdot g=0,\\
(D_x^{3}-{3}a D_{t_{2}}D_x)f\cdot g=0, \qquad  a =\pm1.
\end{array}\right.
\end{equation}
In a similar way, by analysing the second equation of the Eqs.  (\ref{eq:ccps}), we have
\begin{equation}\label{eq:ffps}
\begin{array}{ll}
P_{2}=[(D_{x}D_{t_{4}}-{4}D_{x}^{3}D_{t_{2}})f\cdot f]g^{2}-f^{2}[((D_{x}D_{t_{4}}-{4}D_{x}^{3}D_{t_{2}})g\cdot g]\\
\quad=[(D_{x}D_{t_{4}}f\cdot f)g^{2}-f^{2}(D_{x}D_{t_{4}}g\cdot g)]-{4}[(D_{x}^{3}D_{t_{2}}f\cdot f)g^{2}-f^{2}(D_{x}^{3}D_{t_{2}}g\cdot g)]\\
\quad={2}D_x(D_{t_{4}}f\cdot g)(g f)-{8}D_{t_{2}}(D_x^{3}f\cdot g)(g f)-{24}D_x(D_xD_{t_{2}}f\cdot g)(D_xg\cdot f).\\
\end{array}
\end{equation}
By now, we consider the following property of the $D$ operator,
\begin{equation}\label{eq:ddsps}
\begin{array}{ll}
    \exp(\varepsilon D_x)(\exp(\delta_{1}D_x+\delta_{2}D_{t_{2}})f\cdot g)(\exp(\delta_{1}D_x-\delta_{2}D_{t_{2}})f\cdot g)\\
   \qquad=\exp(\delta_{2}D_{t_{2}})(\exp(\delta_{1}D_x+\varepsilon D_x)f\cdot g)(\exp(\delta_{1}D_x-\varepsilon D_x)f\cdot g).
\end{array}
\end{equation}
Consider the coefficient of the term $\varepsilon\delta_{2}\delta_{1}^{2}$ in Eq.(\ref{eq:ddsps}),
\begin{equation}\label{eq:dddps}
\begin{array}{ll}
\textrm{LHS}={2}D_x[\frac{3}{3!}(D_x^{2}D_{t_{2}} f\cdot g)(f g)-\frac{1}{2!}(D_x^{2}f\cdot g)(D_{t_{2}}f\cdot g)+\frac{2}{2!}(D_xD_{t_{2}}f\cdot g)(D_x f\cdot g)]\\
\qquad={2}D_x[\frac{1}{2}(D_x^{2}D_{t_{2}} f\cdot g)(f g)+\frac{1}{2}(D_x^{2}f\cdot g)(D_{t_{3}}g\cdot f)-(D_xD_{t_{2}}f\cdot g)(D_x g\cdot f)],
\end{array}
\end{equation}
\begin{equation}\label{eq:ddddps}
\begin{array}{ll}
\textrm{RHS}={2}D_{t_{2}}[\frac{3}{3!}(D_x^{3} f\cdot g)(f g)-\frac{1}{2!}(D_x^{2}f\cdot g)(D_x f\cdot g)+\frac{2}{2!}(D_x^{2}f\cdot g)(D_x f\cdot g)]\\
\qquad={2}D_{t_{2}}[\frac{1}{2}(D_x^{3} f\cdot g)(f g)-\frac{1}{2}(D_x^{2}f\cdot g)(D_x g\cdot f)].
\end{array}
\end{equation}
From the first equation of the Eq. (\ref{eq:ddps}) $(D_{t_{2}}+a D_x^{2})f\cdot g=0$, we can get
\begin{equation}\label{eq:dddddps}
    D_x(D_x^{2}f\cdot g)(D_{t_{2}}f\cdot g)=0.
\end{equation}
From Eqs. (\ref{eq:dddps}), (\ref{eq:ddddps}), (\ref{eq:dddddps}), we have
\ben\label{eq:eeps}
    D_x(D_xD_{t_{2}}f\cdot g)(D_x g\cdot f)&=&\frac{1}{2}D_x(D_x^{2}D_{t_{2}}f\cdot g)(gf)+
\frac{1}{2}D_{t_{2}}(D_x^{2}f\cdot g)(D_xg\cdot f)-\nonumber\\
  & &  \frac{1}{2}D_{t_{2}}(D_x^{3}f\cdot g)(g f).
\een
After substituting Eq. (\ref{eq:eeps}) into Eq. (\ref{eq:ffps}), we obtain
\ben\label{eq:fffps}
P_{2}&=&{2}D_x(D_{t_{4}}f\cdot g)(g f)-{8}D_{t_{2}}(D_x^{3}f\cdot g)(g f)-{12}D_x(D_x^{2}D_{t_{2}}f\cdot g)(g f)-\nonumber\\
&&{12}D_{t_{2}}(D_x^{2}f\cdot g)(D_xg\cdot f)+{12}D_{t_{2}}(D_x^{3}f\cdot g)(g f)\nonumber\\
&=&{2}D_x[(D_{t_{4}}-{6}D_x^{2}D_{t_{2}})f\cdot g](g f)+{4}D_{t_{2}}(D_x^{3}f\cdot g)(g f)+\nonumber\\
&&{12}aD_{t_{2}}(D_{t_{2}}f\cdot g)(D_xf\cdot g).
\een
According to the realtions
\ben
D_{t_{2}}(D_{t_{2}}f\cdot g)(D_xf\cdot g)=D_x(D_{t_{2}}^{2}f\cdot g)(g f)-
D_{t_{2}}(D_xD_{t_{2}}f\cdot g)(g f),\\
(D_x^{3}-{3}a D_{t_{2}}D_x)f\cdot g=0,
\een
Eq. (\ref{eq:fffps}) can be rewritten as
\ben
P_{2}&=&{2}D_x[(D_{t_{4}}-{6}D_x^{2}D_{t_{2}})f\cdot g](g f)+{12}aD_{t_{2}}(D_{t_{2}}D_xf\cdot g)(g f)+\nonumber\\
&&{12}aD_x(D_{t_{2}}^{2}f\cdot g)(g f)-{12}aD_{t_{2}}(D_xD_{t_{2}}f\cdot g)(g f)\nonumber\\
&=&{2}D_x[(D_{t_{4}}-{6}D_x^{2}D_{t_{2}}+{6}aD_{t_{2}}^{2})f\cdot g](g f).
\een
As $P_{2}=0$, we can get
\begin{equation}\label{eq:ffffps}
    (D_{t_{4}}-{6}D_x^{2}D_{t_{2}}+{6}aD_{t_{2}}^{2})f\cdot g=0.
\end{equation}
By combining Eqs. (\ref{eq:ddps}) and (\ref{eq:ffffps}), one have
\begin{equation}\label{eq:bou}
\left\{\begin{array}{ll}
(D_{t_{2}}+a D_x^{2})f\cdot g=0,\\
(D_x^{3}-{3}a D_{t_{2}}D_x)f\cdot g=0 ,\qquad \qquad a=\pm1.\\
(D_{t_{4}}-{6}D_x^{2}D_{t_{2}}+{6}aD_{t_{2}}^{2})f\cdot g=0,
\end{array}\right.
\end{equation}

 Therefore, we obtain the bilinear B\"{a}cklund transformation of the Bousinessq equation. One can know the relations of the solutions of the Bousinessq equations or get other solutions from this transformation.
\section{The Lax pair of the Boussinesq equation }
In this section, we  will obtain the Lax pair of the Boussinesq equation through their B\"{a}cklund trabsformation in \S 3.
 To begin with, we transform the bilinear B\"{a}cklund transformation
\begin{equation}\label{eq:bbou}
\left\{\begin{array}{ll}
(D_{t_{2}}+a D_x^{2})f\cdot g=0,\\
(D_x^{3}-{3}a D_{t_{2}}D_x)f\cdot g=0, \qquad \qquad a=\pm1.\\
(D_{t_{4}}-{6}D_x^{2}D_{t_{2}}+{6}aD_{t_{2}}^{2})f\cdot g=0,
\end{array}\right.
\end{equation}
to nonlinear form.
 Let $a=1$, and
\begin{equation}\label{eq:aaeq}
    \psi=\frac{f}{g},\quad u={2}(\ln g)_{xx},\quad w=\frac{g_{t_{2}}}{g}\quad p=\frac{g_{t_{2}t_{2}}}{g}.
\end{equation}
According to the relations
\begin{equation}\label{eq:bbeq}
\begin{array}{ll}
\frac{D_{t_{2}}f\cdot g}{g^{2}}=\psi_{t_{2}},\\
\frac{D_{t_{4}}f\cdot g}{g^{2}}=\psi_{t_{4}},\\
\frac{D_xD_{t_{2}}f\cdot g}{g^{2}}=\psi_{xt_{2}}+{2}w_x\psi,\\
\frac{D_x^{2}f\cdot g}{g^{2}}=\psi_{xx}+u\psi,\\
\frac{D_x^{3}f\cdot g}{g^{2}}=\psi_{xxx}+{3}u\psi_x,\\
\frac{D_x^{2}D_{t_{2}}f\cdot g}{g^{2}}=\psi_{xxt_{2}}+u\psi_{t_{2}}+{4}w_x\psi_x,\\
\frac{D_{t_{2}}^{2}f\cdot g}{g^{2}}=\psi_{t_{2}t_{2}}+{2}(p-w^{2})\psi,
\end{array}
\end{equation}
one can get the nonlinear differential equations by eliminating $t_{2}$ from Eqs. (\ref{eq:bbou})
\begin{equation}\label{eq:bbbou}
\left\{\begin{array}{ll}
\psi_{xxx}+{3}u\psi_x-{3}\psi_{xt_{2}}-{6}w_x\psi=0,\\
\psi_{t_{2}}+\psi_{xx}+u\psi_x=0,\\
\psi_{t_{4}}-{6}\psi_{xxt_{2}}-{6}u\psi_{t_{2}}-{24}w_x\psi_x+{6}\psi{t_{2}t_{2}}+{12}(p-w^{2})\psi=0.
\end{array}\right.
\end{equation}
In order to get the Lax pair of the above equations, we suppose
\begin{equation}\label{eq:ccbou}
\begin{array}{ll}
\psi_{1}=\psi_x,\quad\psi_{2}=\psi_{xx},\quad\phi=\psi_{t_{2}}\\
\Psi=(\phi,\psi,\psi_{1},\psi_{2}).
\end{array}
\end{equation}
The Lax matrix $U, V$ or Lax pair will satisfy the relations
\begin{equation}\label{eq:ddbou}
\left\{\begin{array}{ll}
\Psi_x=U\Psi,\\
\Psi_{t_{4}}=V\Psi.
\end{array}\right.
\end{equation}

One can get that
\begin{equation}\label{eq:eebou}
\begin{array}{ll}
\phi_x=-\psi_{xxx}-u_x\psi-u\psi_x\\
\quad=\frac{3}{2}u\psi_{x}+\frac{3}{4}(u_x-{2}w_x)\psi-u_x\psi-u\psi_x\\
\quad=\frac{1}{2}u\psi_x-\frac{1}{4}(u_x+{6}w_x)\psi\\
\quad=\frac{1}{2}u\psi_{1}-\frac{1}{4}(u_x+{6}w_x)\psi,\\
(\psi_{2})_x=-\frac{3}{2}u\psi_{x}-\frac{3}{4}(u_x-{2}w_x)\psi\\
\quad=-\frac{3}{2}u\psi_{1}-\frac{3}{4}(u_x-{2}w_x)\psi.
\end{array}
\end{equation}
Therefore, one can obtain the Lax matrix U in Eq.(\ref{eq:ddbou}) as
\begin{displaymath}
\mathbf{U}=
\left(\begin{array}{cccc}
  0 &\frac{1}{4}(u_x+{6}w_x) & \frac{1}{2}u & 0  \\
  0 & 0 & 1 & 0 \\
  0 & 0 & 0 & 1 \\
  0 & -\frac{3}{4}(u_x-{2}w_x) & -\frac{3}{2}u & 0
\end{array}\right).
\end{displaymath}
In a similar way , we can get the following equation from the second equation of the Eqs. (\ref{eq:bbbou})
\begin{equation}\label{eq:tj}
    \psi_{xxt_{2}}=-\psi_{t_{2}t_{2}}-{2}w_{xx}\psi-u\psi_{t_{2}}.
\end{equation}
We can also get the equation from the first equation of the Eqs. (\ref{eq:bbbou})
\begin{equation}\label{eq:ttj}
\begin{array}{ll}
    \psi_{xxt_{2}}=\frac{1}{3}\psi_{xxxx}+u_x\psi_x+u\psi_{xx}-{2}w_{xx}\psi-{2}w_x\psi_x\\
    \qquad=\frac{1}{3}(-\frac{3}{2}u\psi_x-\frac{3}{4}(u_x-{2}w_x)\psi)_x+u\psi_{xx}+(u_x-{2}w_x)\psi_x-{2}w_{xx}\psi\\
    \qquad=\frac{1}{2}u\psi_{xx}+\frac{1}{4}(u_x-{6}w_x)\psi_x-\frac{1}{4}(u_{xx}+{6}w_{xx})\psi,
\end{array}
\end{equation}
Combining Eqs.(\ref{eq:tj}), (\ref{eq:ttj}), we have
\begin{equation}
    \psi_{t_{2}t_{2}}=-\frac{1}{2}u\psi_{xx}-\frac{1}{4}(u_x-{6}w_x)\psi_x+\frac{1}{4}(u_{xx}-{2}w_{xx})\psi-u\psi_{t_{2}}.
\end{equation}

Taking advantage of  Eqs. (\ref{eq:bbbou}) and the above equations, we get the following more relations
\begin{equation*}
\begin{array}{ll}
\psi_{t_{4}}={24}w_x\psi_x-{12}(w_{xx}+(p-w^{2}))\psi-{12}\psi_{t_{2}t_{2}}\\
\quad={24}w_x\psi_x-{12}(w_{xx}+(p-w^{2}))\psi-{6}u\psi_{xx}+{3}(u_x-{6}w_x)\psi_x-{3}(u_{xx}-{2}w_{xx})\psi+{12}u\psi_{t_{2}}\\
\quad={6}u\psi_{xx}+{3}(u_x+{2}w_x)\psi_x-{3}(u_{xx}+{2}w_{xx}+{4}(p-w^{2}))\psi+{12}u\psi_{t_{2}}\\
\quad={6}u\psi_{2}+{3}(u_x+{2}w_x)\psi_{1}-{3}(u_{xx}+{2}w_{xx}+{4}(p-w^{2}))\psi+{12}u\phi,\\
(\psi_{1})_{t_{4}}={6}u_x\psi_{xx}+{6}\psi_{xxx}+{3}(u_xx+{2}w_{xx})\psi_x+{3}(u_x+2w_x)\psi_{xx}-{3}(u_{xxx}+{2}w_{xxx}+{4}(p-w^{2}))\psi\\
\qquad-{3}(u_{xx}+{2}w_{xx}+{4}(p-w^{2}))\psi_x+{12}u_x\psi_{t_{2}}+{12}u\psi_{xt_{2}}\\
\quad={6}u\psi_{xx}+{6}u(-\frac{3}{2}u\psi_x-\frac{3}{4}(u_x-{2}w_x)\psi)+{3}(u_xx+{2}w_{xx})\psi_x+{3}(u_x+2w_x)\psi_{xx}\\
\qquad-{3}(u_{xxx}+{2}w_{xxx}+{4}(p-w^{2}))\psi-{3}(u_{xx}+{2}w_{xx}+{4}(p-w^{2}))\psi_x\\
\qquad+{12}u_x\psi_{t_{2}}+{12}u(\frac{1}{2}u\psi_x-\frac{1}{4}(u_x+{6}w_x))\psi\\
\quad={3}({3}u_x+{2}w_x)\psi_{xx}-{3}(u^{2}+{4}(p-w^{2}))\psi_x-{3}(u_{xxx}+{2}w_{xxx}+{4}(p-w^{2})_x+\frac{5}{2}uu_x+{3}uw_x)\psi\\
\qquad+{12}u_x\psi_{t_{2}}\\
\quad={3}({3}u_x+{2}w_x)\psi_{2}-{3}(u^{2}+{4}(p-w^{2}))\psi_{1}-{3}(u_{xxx}+{2}w_{xxx}+{4}(p-w^{2})_x+\frac{5}{2}uu_x+{3}uw_x)\psi\\
\qquad+{12}u_x\phi,
\end{array}
\end{equation*}
\begin{equation*}
\begin{array}{ll}
(\psi_{2})_{t_{4}}=({9}u_{xx}+{6}w_{xx})\psi_{xx}+({9}u_x+{6}w_x)\psi_{xxx}-({6}uu_x+{12}(p-w^{2}))\psi_x\\
\qquad-({3}u^{2}+{12}(p-w^{2})\psi_{xx}-({3}u_{xxxx}++{6}w_{xxxx}+{12}(p-w^{2})_{xx}+\frac{15}{2}(uu_x)_x+{9}(uw_x)_x)\psi\\
\qquad-({3}u_{xxx}+{6}w_{xxx}+{12}(p-w^{2})_x+\frac{15}{2}uu_x+{9}uw_x)\psi_x+{12}u_{xx}\psi_{t_{2}}+{12}u_x\psi_{xt_{2}}\\
\quad=({9}u_{xx}+{6}w_{xx})\psi_{xx}+({9}u_x+{6}w_x)(-\frac{3}{2}u\psi_x-\frac{3}{4}(u_x-{2}w_x)\psi)-({6}uu_x+{12}(p-w^{2})_x)\psi_x\\
\qquad-({3}u^{2}+{12}(p-w^{2}))\psi_{xx}-({3}u_{xxxx}+{6}w_{xxxx}+{12}(p-w^{2})_{xx}+\frac{15}{2}u_x^{2}+\frac{15}{2}uu_{xx}+{9}u_xw_x\\
\qquad+{9}uw_{xx})\psi-({3}u_{xxx}+{6}w_{xxx}+{12}(p-w^{2})_x+\frac{15}{2}uu_x+{9}uw_x)\psi_x+{12}u_{xx}\psi_{t_{2}}\\
\qquad+{12}u_x(\frac{1}{2}u\psi_x-\frac{1}{4}(u_x+{6}w_x)\psi)\\
\quad={3}({3}u_{xx}+{2}w_{xx}-u^{2}-{4}(p-w^{2}))\psi_{xx}-{3}(u_{xxx}+{2}w_{xxx}+{8}(p-w^{2})_x+{7}uu_x+{6uw_x})\psi_x\\
\qquad-{3}(u_{xxxx}+{2}w_{xxxx}+{4}(p_{xx}-{2}ww_{xx})-{11}w_x^{2}+\frac{23}{4}u_x^{2}+\frac{5}{2}uu_{xx}+{6}u_xw_x+{3}uw_{xx})\psi\\
\qquad+{12}u_{xx}\psi_{t_{2}}\\
\quad={3}({3}u_{xx}+{2}w_{xx}-u^{2}-{4}(p-w^{2}))\psi_{2}-{3}(u_{xxx}+{2}w_{xxx}+{8}(p-w^{2})_x+{7}uu_x+{6uw_x})\psi_{1}\\
\qquad-{3}(u_{xxxx}+{2}w_{xxxx}+{4}(p_{xx}-{2}ww_{xx})-{11}w_x^{2}+\frac{23}{4}u_x^{2}+\frac{5}{2}uu_{xx}+{6}u_xw_x+{3}uw_{xx})\psi\\
\qquad+{12}u_{xx}\phi,\\
\phi_{t_{4}}=-\psi_{xxt_{4}}-u_{t_{4}}\psi-u\psi_{t_{4}}\\
\quad=-{3}({3}u_{xx}+{2}w_{xx}-u^{2}-{4}(p-w^{2}))\psi_{2}+{3}(u_{xxx}+{2}w_{xxx}+{8}(p-w^{2})_x+{7}uu_x+{6uw_x})\psi_{1}\\
\qquad+{3}(u_{xxxx}+{2}w_{xxxx}+{4}(p_{xx}-{2}ww_{xx})-{11}w_x^{2}+\frac{23}{4}u_x^{2}+\frac{5}{2}uu_{xx}+{6}u_xw_x+{3}uw_{xx})\psi\\
\qquad-{12}u_{xx}\phi-_{t_{4}}\psi-u({6}u\psi_{xx}+{3}(u_x+{2}w_x)\psi_x-{3}(u_{xx}+{2}w_{xx}+{4}(p-w^{2}))\psi+{12}u\psi_{t_{2}})\\
\quad=-{3}({3}u_{xx}+{2}w_{xx}+u^{2}-{4}(p-w^{2}))\psi_{xx}+{3}(u_{xxx}+{2}w_{xxx}+{8}(p-w^{2})_x+{6}uu_x+{4}uw_x)\psi_x\\
\qquad+{3}(u_{xxxx}+{2}w_{xxxx}+{4}(p_{xx}-{2}ww_{xx})-{11}w_x^{2}+\frac{13}{4}u_x^{2}+\frac{7}{2}uu_{xx}+{6}u_xw_x-\frac{1}{3}u_{t_{4}}\\
\qquad+{4}u(p-w^{2})+{5}uw_{xx})\psi-{12}(u_{xx}+u^{2})\psi_{t_{2}}\\
\quad=-{3}({3}u_{xx}+{2}w_{xx}+u^{2}-{4}(p-w^{2}))\psi_{2}+{3}(u_{xxx}+{2}w_{xxx}+{8}(p-w^{2})_x+{6}uu_x+{4}uw_x)\psi_{1}\\
\qquad+{3}(u_{xxxx}+{2}w_{xxxx}+{4}(p_{xx}-{2}ww_{xx})-{11}w_x^{2}+\frac{13}{4}u_x^{2}+\frac{7}{2}uu_{xx}+{6}u_xw_x-\frac{1}{3}u_{t_{4}}\\
\qquad+{4}u(p-w^{2})+{5}uw_{xx})\psi-{12}(u_{xx}+u^{2})\phi.
\end{array}
\end{equation*}
From which, we can get another Lax matrix V as,
\begin{equation}
 \mathbf{V}=(v_{i j})\qquad i,j=1,2,3,4.
\end{equation}
\begin{equation*}
\begin{array}{ll}
where\\
v_{11}=-{12}(u_{xx}+u^{2}),\\
v_{12}={3}(u_{xxxx}+{2}w_{xxxx}+{4}(p_{xx}-{2}ww_{xx})-{11}w_x^{2}+\frac{13}{4}u_x^{2}+\frac{7}{2}uu_{xx}+{6}u_xw_x-\frac{1}{3}u_{t_{4}}\\
\qquad\qquad\qquad+{4}u(p-w^{2})+{5}uw_{xx}),\\
v_{13}={3}(u_{xxx}+{2}w_{xxx}+{8}(p-w^{2})_x+{6}uu_x+{4}uw_x),\\
v_{14}=-{3}({3}u_{xx}+{2}w_{xx}+u^{2}-{4}(p-w^{2})) ,
\end{array}
\end{equation*}
\begin{equation*}
\begin{array}{ll}
v_{21}={12}u,\\
v_{22}=-{3}(u_{xx}+{2}w_{xx}+{4}(p-w^{2})), \\
v_{23}={3}(u_x+{2}w_x),\\
v_{24}={6}u,\\
v_{31}={12}u_x,\\
v_{32}=-{3}(u_{xxx}+{2}w_{xxx}+{4}(p-w^{2})_x+\frac{5}{2}uu_x+{3}uw_x),\\
v_{33}=-{3}(u^{2}+{4}(p-w^{2})),\\
v_{34}={3}({3}u_x+{2}w_x),\\
v_{41}={12}u_{xx},\\
v_{42}=-{3}(u_{xxxx}+{2}w_{xxxx}+{4}(p_{xx}-{2}ww_{xx})-{11}w_x^{2}+\frac{23}{4}u_x^{2}+\frac{5}{2}uu_{xx}+{6}u_xw_x+{3}uw_{xx}),\\
v_{43}=-{3}(u_{xxx}+{2}w_{xxx}+{8}(p-w^{2})_x+{7}uu_x+{6uw_x}),\\
v_{44}={3}({3}u_{xx}+{2}w_{xx}-u^{2}-{4}(p-w^{2})).
\end{array}
\end{equation*}

It can be verified that the Lax pair satisfy the zero curvature condition. The equations by eliminating $t_{2}$ form the Boussinesq equations is integrable.
\section{Conclutions}
In this paper, we have studied the Boussinesq  equation from the viewpoint of Hirota¡¯s bilinear method. We have derived the bilinear Backlunsd transformation and Lax pair of this equation. It is obvious that one can also use this method to discuss other bilinear differential equations or supersymmetris nonlinear differential equations.

{\bf Acknowledgments:}
The project was supported by National Natural
Science Foundation of China (Grant No. 10801045).

\end{document}